\documentclass{article}
\usepackage{spconf,amsmath,graphicx,hyperref}

\usepackage{spconf}

\usepackage{cite}
\usepackage{amsmath,amssymb,amsfonts}
\usepackage{dsfont}
\usepackage{algorithmic}
\usepackage{graphicx}
\usepackage{textcomp}
\usepackage{wrapfig}
\usepackage{kotex}
\usepackage{xcolor}
\usepackage[T1]{fontenc}
\usepackage{tabularray}
\usepackage{multirow}
\usepackage{multicol}
\usepackage{booktabs}
\usepackage{amsmath}
\usepackage{makecell}
\usepackage{orcidlink}

\usepackage{float}
\usepackage{tabularx}
\usepackage{subcaption}
\usepackage{colortbl}
\usepackage{pgf}
\usepackage{hyperref}

\newif\ifdebug

\newcounter{commentcounter}
\newcounter{unknowncounter}
\newcounter{needchartcounter}

\newcommand{\needchart}[1]{\protect\stepcounter{needchartcounter}\textcolor{blue}{Fig. XX}}

\newcommand{\debug}[1]{\ifdebug#1\fi}

\usepackage{xparse}

\newcounter{mycounter}

\newcommand{\heatmap}[3]{%
  \pgfmathparse{int((#1-#2)/(#3-#2)*55)}%
  \edef\temp{\noexpand\cellcolor{red!\pgfmathresult}}%
  \temp #1%
}

\title{MOUSSE: Multilingual Overlapped Speech Separation with Predictive Uncertainty Estimation}

\name{Su Yeon Shim\,\orcidlink{0009-0002-9518-0827}$^{1}$, Yejin Jeon\,\orcidlink{0009-0007-8802-2675}$^{2,3}$, David Ifeoluwa Adelani\,\orcidlink{0000-0002-0193-2083}$^{2,3,4}$, Young-Joo Suh\,\orcidlink{0000-0001-7208-1709}$^{5,6}$}
\address{$^1$KT Corporation
$^2$Mila - Quebec AI Institute
$^3$McGill University \\
$^4$Canada CIFAR AI Chair
$^5$Graduate School of Artificial Intelligence, POSTECH \\
$^6$Department of Computer Science and Engineering, POSTECH}

\begin{document}

\maketitle

\begin{abstract}
Recent generative speech separation methods have shown promising performance, yet mixtures of overlapping speech in different languages remain underexplored. To this end, we first curate MuSSeD, an English--Korean benchmark comprising two- to four- speaker mixtures across varying overlap ratios. We further propose MOUSSE, a flow matching framework that incorporates boundary-aware training and frame-wise predictive uncertainty through a permutation-invariant Gaussian negative log-likelihood. The predicted scale is further used to adapt classifier-free guidance during sampling. MOUSSE achieves 13.18~dB SI-SDRi and 0.175~ESTOIi, outperforming discriminative and flow matching-based separation baselines, with a 1.58~dB SI-SDRi improvement over the strongest model.
\end{abstract}


\section{Introduction}
\label{sec:introduction}

Speech separation, which aims to recover individual speakers' signals from overlapping utterances, is central to the cocktail-party problem. Despite substantial progress, existing systems are developed predominantly in monolingual settings, especially English, with fixed speaker counts~\cite{borsdorf2021globalphone, wang2022disentangling}. Real-world conversations, by contrast, involve varying speaker counts, degrees of temporal overlap, and multiple languages, with cross-lingual differences in phonetic and intonational patterns introducing additional variability. A further challenge specific to multilingual settings is disentangling language identity from individual-speaker separation. Prior work on blind language separation (BLS)~\cite{borsdorf2022blind} emphasizes language-based disentanglement and generalization to unseen languages, but leaves open how reliably individual speakers are recovered as language composition varies. Therefore, systematic evaluations across languages, speaker counts, and overlap ratios are important in order to assess the robustness of multilingual speech separation.

Prior architectures have made significant progress in improving monolingual speech separation. Discriminative models such as Conv-TasNet~\cite{luo2019conv} introduced an end-to-end time-domain approach, while RE-SepFormer~\cite{della2024resource} improves Transformer efficiency through compact representations. TIGER~\cite{xu2025tiger} and SR-CorrNet~\cite{shin2026asymmetric} further exploit time-frequency modeling and spatio-spectro-temporal correlations via an asymmetric encoder-decoder, respectively. Despite their effectiveness, these methods typically produce a single set of source estimates through direct inference, offering limited opportunity to iteratively refine uncertain source components.

Generative approaches offer a complementary approach, casting separation as an iterative reconstruction process. DiffSep~\cite{scheibler2023diffusion} formulates separation via a score-matching diffusion process, while SpeechFlow~\cite{liu2024generative} employs flow matching within a unified framework spanning separation, enhancement, and text-to-speech. UA-Flow~\cite{han2026flow} extends this line of work by jointly predicting a heteroscedastic uncertainty estimate alongside the velocity field in vision tasks. While classifier-free guidance (CFG) can strengthen conditional separation by amplifying the influence of the mixture during sampling, standard CFG~\cite{ho2022classifier} uses a shared guidance strength across sequence positions, without accounting for frame-wise variation in speaker activity and temporal overlap. This motivates adapting the guidance strength across frames.

To address these gaps, we propose MOUSSE, a flow matching framework for multilingual speech separation with predictive uncertainty. Our contributions are threefold: (1) we construct MuSSeD, an English–Korean speech separation benchmark spanning two to four speakers, English-only, Korean-only, and mixed-language compositions, and multiple overlap ratios, (2) we integrate a frame-wise predictive uncertainty estimator into a flow matching separator for uncertainty-aware training and sampling, and (3) we introduce boundary-aware weighting that accounts for speaker activity and assigns larger weights to frames near activity transitions. Extensive experiments show that our method outperforms discriminative and generative baselines on MuSSeD. The code and dataset will be made publicly available.\footnote{\url{https://github.com/SHIM-SUYEON/MouSSe.git}}\footnote{\url{https://huggingface.co/datasets/syshim/MuSSeD-EnKo}}

\section{Method}
\label{sec:method}

\subsection{Speech Separation Dataset Curation}

For bilingual speech separation, we curate the MuSSeD dataset by mixing English speech from LibriTTS-R~\cite{koizumi2023libritts} and Korean speech from the Multi-Speaker Speech Synthesis corpus~\cite{aihub2021}. The resulting dataset contains 37,800 mixtures from 514 speakers, with a male-to-female ratio of approximately 55:45. For each speaker count, the training, validation, and test sets contain 10,000, 1,300, and 1,300 mixtures, respectively. Speakers are mutually disjoint across the three splits to prevent speaker leakage between training and evaluation. Each mixture lasts 3--15 seconds and contains two to four speakers. We generate English-only, Korean-only, and mixed-language mixtures with overlap ratios of 0\%, 25\%, 50\%, 75\%, and 100\%. Note that the overlap ratio is defined for each pair of adjacent speakers as the overlap duration relative to the duration of the shorter utterance. For mixtures with more than two speakers, speakers are placed sequentially such that the specified ratio is applied to each adjacent pair.

\subsection{Uncertainty-Aware Training}
\label{sec:method-training}

We introduce boundary-aware weighting for the separation and uncertainty objectives with $S$ output slots (i.e., separated audio sources). Weighting focuses the objectives on frames, in which the target speaker is active, while assigning additional weight to frames near activity transitions. The separator follows the SpeechFlow backbone~\cite{liu2024generative}, which adapts a pre-trained flow matching model through task-specific conditioning for speech separation, and is trained as a single model on two- to four-speaker mixtures.

\begin{figure}[t]
    \centering
    \includegraphics[width=0.9\linewidth]{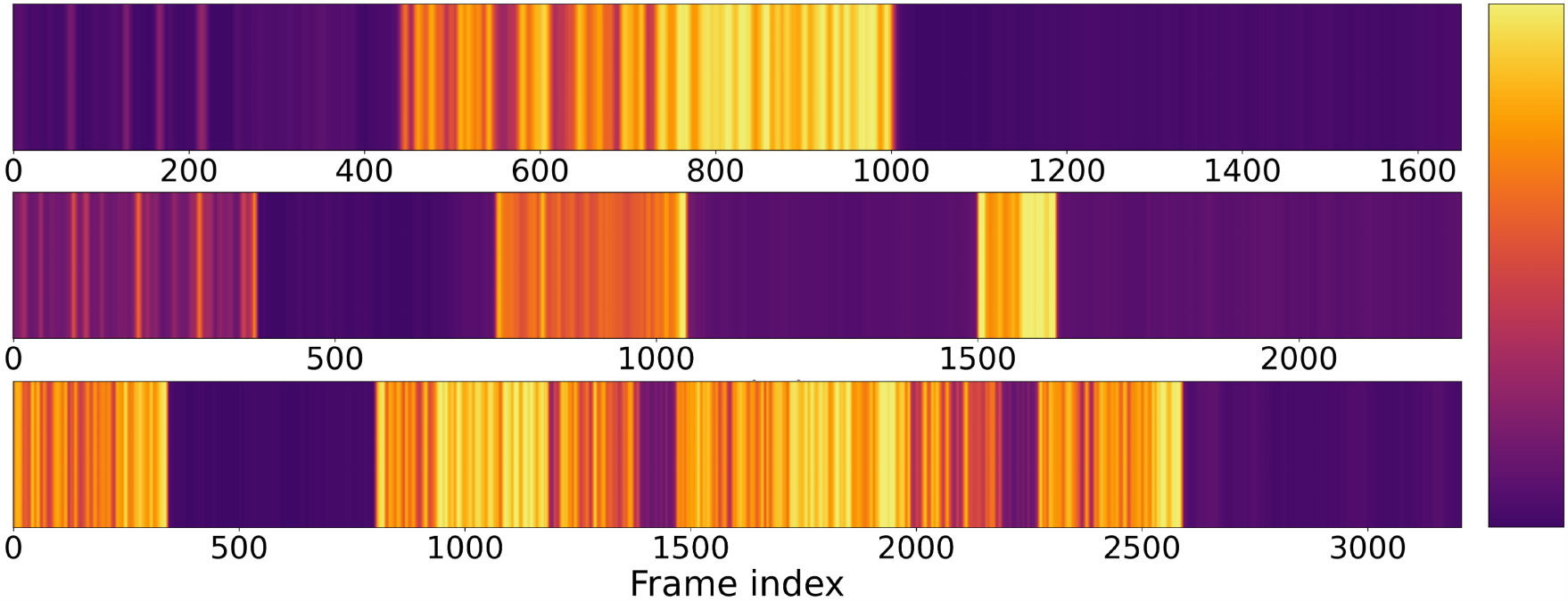}
    \caption{Boundary-aware weight maps for two- to four-speaker mixtures, which illustrate frames near activity boundaries.}
    \label{fig:entropy_map}
\end{figure}

Formally, for target speaker $k$ at frame $t$, we obtain a binary activity indicator $a_{k,t}\in\{0,1\}$ from the target mel-spectrogram. Activity boundaries are identified from consecutive state changes, and a window of $K=15$ frames around each boundary is marked as the boundary region. Training weight is defined as:

\begin{equation}
    w_{k,t}=a_{k,t}(1+b_{k,t}),
    \label{eq:boundary_weight}
\end{equation}
where $b_{k,t}\in\{0,1\}$ denotes the boundary indicator, set to 1 within the boundary neighborhood and 0 otherwise. Active frames near activity boundaries are therefore assigned twice the weight of steady active frames. Figure~\ref{fig:entropy_map} visualizes the resulting boundary-aware weight map.

Resulting boundary-aware weights are incorporated into the separation objective. The pairwise weighted mean squared error, using $w_{k,t}$ from Equation~\ref{eq:boundary_weight}, is given by
\begin{equation} 
    \ell_{\mathrm{MSE}}(s,k) = \frac{1}{W_k D} \sum_t w_{k,t} \left\| \hat{v}_{s}(t)-v^{*}_{k}(t) \right\|_2^2,
    \label{eq:weighted_mse} 
\end{equation}
where $\hat {v}_s(t)$ denotes the predicted velocity for output slot $s$ at frame $t$, and $v_k^*(t)$ represents the target flow velocity for speaker $k$. $W_k=\sum_t w_{k,t}$ is the total weight for speaker $k$ and $D$ is the number of mel dimensions. The velocity describes the flow from the noise distribution toward the target distribution. To account for speaker permutation ambiguity, permutation-invariant training (PIT) selects the assignment $\pi_{\mathrm{MSE}}$ from the losses, resulting in the separation objective:

\begin{equation}
    \mathcal{L}_{\mathrm{MSE}} = \frac{1}{S} \sum_{s=1}^{S} \ell_{\mathrm{MSE}}\bigl(s,\pi_{\mathrm{MSE}}(s)\bigr).
    \label{eq:pit_mse}
\end{equation}

Furthermore, in order to identify regions where the model struggles to disentangle overlapping speech, we propose a predictive uncertainty estimator. The estimator takes the observed mixture $x$, flow time $\tau$, and current flow state $z_\tau$. It uses sinusoidal and convolutional positional embeddings followed by 12 Transformer blocks with hidden dimension 1024 and 8 attention heads. 
This uncertainty estimator then predicts $\log \sigma_s$, a representation of the scale, at each frame and mel dimension for the output slot, providing a smooth parameterization of the positive scale required by the Gaussian likelihood. The corresponding scale $\sigma_s$ is obtained by exponentiation, and the pairwise Gaussian negative log-likelihood is defined as:

\begin{equation}
    \ell_{\mathrm{NLL}}(s,k) = \frac{1}{W_kD}\sum_{t,d} w_{k,t} \left[ \frac{r_{s,k}^{2}}{2\sigma_s^{2}} + \log\sigma_s \right],
    \label{eq:weighted_nll}
\end{equation}
where $r_{s,k}$ is the residual between the detached separator prediction $\hat{v}_s$ and the target velocity $v_k^*$, and $\sigma_s$ is the corresponding predicted scale, both evaluated at each frame and mel dimension. PIT selects $\pi_{\mathrm{NLL}}$ from the corresponding pairwise likelihood losses while keeping the separator prediction and predicted scale paired within each output slot. The uncertainty objective and the total training objective are

\begin{figure*}[!t]
    \centering
    \includegraphics[width=0.85\textwidth]{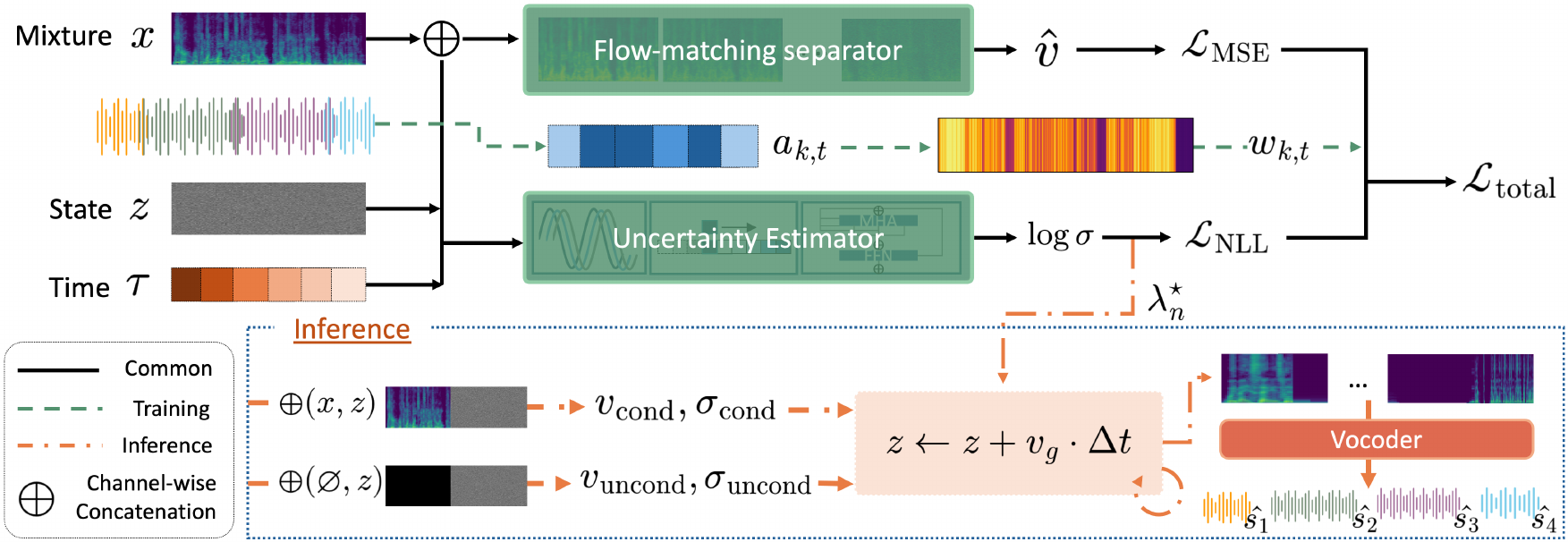}
    \caption{Overview of MOUSSE with boundary-aware training and adaptive uncertainty-guided sampling.}
    \label{fig:overview}
\end{figure*} 

\begin{equation}
    \mathcal{L}_{\mathrm{NLL}}=\frac{1}{S}\sum_{s=1}^{S}\ell_{\mathrm{NLL}}\bigl(s,\pi_\mathrm{NLL}(s)\bigr),
    \label{eq:pit_nll}
\end{equation}

\begin{equation}
    \mathcal{L}_{\mathrm{total}} = \mathcal{L}_{\mathrm{MSE}} + \beta \cdot \mathcal{L}_{\mathrm{NLL}},
    \label{eq:total_loss}
\end{equation}
with $\beta=0.2$. Note that the two objectives share the same boundary-aware weights, assigning larger weights to frames near speaker activity boundaries.

\subsection{Adaptive Uncertainty-Guided Sampling}

Recall that in Section \S\ref{sec:method-training}, the uncertainty estimator predicts a scale $\sigma_{s}$ for each position $n$ and mel dimension $d$. To account for variations in the predicted scale across sequence positions, we adapt the classifier-free guidance (CFG) strength accordingly during sampling. Thus, the mean scale across mel dimensions defines the adaptive guidance strength:

\begin{equation}
    \lambda^\star_{s, n} = \min \Big( \gamma \cdot \frac{1}{D} \sum_d \sigma_s(n,d),\lambda_{\max} \Big),
    \label{eq:adaptive_lambda}
\end{equation}
where $\gamma$ controls the uncertainty scaling, and $\lambda_{\max}=2.0$ bounds the guidance strength. During CFG, the separator produces conditional and unconditional velocity predictions at each frame. The conditional velocity $v_{c,n}$ is predicted using the observed mixture, while the unconditional velocity $v_{\varnothing,n}$ is obtained by replacing the mixture input with zeros. The guided velocity can then be defined as:

\begin{equation}
    v_{g,n}=v_{c,n}+\lambda_n^{\star}\left(v_{c,n}-v_{\varnothing,n}\right).
    \label{eq:adaptive_guidance}
\end{equation}

Moreover, guidance strength increases with predicted scale. Conditioned on the mixture $x$, flow time $\tau$, and current flow state $z_\tau$, the uncertainty estimator updates $\log\sigma_s$ at each frame and mel dimension for output slot $s$. Note that sampling starts from Gaussian noise and proceeds with Euler updates using $\Delta\tau=1/16$. The resulting sequence is split into $S$ separated mel-spectrogram outputs and subsequently converted to waveforms with the BigVSAN~\cite{shibuya2024bigvsan} vocoder.

\section{Experimental Setup}
\label{sec:setup}
MOUSSE is trained for 29000 iterations, and the scaling factor is set to $\gamma=1.0$ as a unit scale so that the predicted uncertainty directly determines the guidance magnitude without additional rescaling. We use AdamW, batch size 4, on a single NVIDIA RTX A6000 GPU.

To assess the proposed MOUSSE model, we juxtapose it against four different baseline models~\cite{luo2019conv, della2024resource, shin2026asymmetric, liu2024generative} for reconstruction fidelity and speech intelligibility using signal-to-distortion ratio improvement (SDRi), scale-invariant signal-to-distortion ratio improvement (SI-SDRi), and extended short-time objective intelligibility improvement (ESTOIi). SDRi is sensitive to signal scaling, SI-SDRi is scale-invariant, and ESTOIi reflects speech intelligibility. For all metrics, improvement is computed relative to the input mixture, using the corresponding clean source as the reference. 

All baseline models are retrained on MuSSeD with the official code and the same test set. Discriminative methods are evaluated directly, whereas predicted mel-spectrograms are decoded using a vocoder for flow matching models. 

\begin{table}[t]
    \centering
    \caption{Overall separation performance on MuSSeD.}
    \label{tab:comparison}
    \begin{tabular}{l|ccc}
        \toprule[1.3pt]
        Model & SDRi& SI-SDRi& ESTOIi \\
        \Xhline{2\arrayrulewidth}
        Conv-TASNET \cite{luo2019conv} & 9.60& 9.72& 0.032 \\
        ReSepformer \cite{della2024resource} & 5.51& 5.56& -0.003 \\
        SR-CorrNet \cite{shin2026asymmetric} & 11.59& 11.60& 0.092 \\
        SpeechFlow \cite{liu2024generative} & 5.04 & 3.65 & -0.072 \\
        \hline
        \hline
        \textbf{MOUSSE (Ours)} & \textbf{13.20} & \textbf{13.18} & \textbf{0.175} \\
        \bottomrule[1.3pt]
    \end{tabular}
\end{table}

\section{Results}
\label{sec:results}

\subsection{Evaluation}
Table~\ref{tab:comparison} reports the overall performance on MuSSeD. Our model achieves 13.20~dB SDRi and 13.18~dB SI-SDRi, outperforming all baseline models. Among the baselines, SR-CorrNet~\cite{shin2026asymmetric} obtains the highest SI-SDRi of 11.60~dB, followed by Conv-TasNet~\cite{luo2019conv} at 9.72~dB, ReSepFormer~\cite{della2024resource} at 5.56~dB, and SpeechFlow~\cite{liu2024generative} at 3.65~dB. MOUSSE therefore improves SI-SDRi by 1.58~dB over the strongest baseline. MOUSSE also achieves the highest intelligibility improvement, reaching 0.175 ESTOIi, while the baselines range from --0.072 to 0.092. Since ESTOIi is measured relative to the input mixture, values near zero indicate little improvement. These results show that MOUSSE improves both signal-level separation and intelligibility.

To evaluate the contribution of each component, we conduct ablation experiments as shown in Table~\ref{tab:ablation}. Removing boundary-aware weighting while retaining the uncertainty objective and adaptive guidance reduces SI-SDRi to 12.10~dB and ESTOIi to 0.164. Additionally, to evaluate the impact of uncertainty modeling, we remove the uncertainty NLL objective. Since adaptive guidance relies on the predicted uncertainty estimates, we use a fixed value with $\lambda=0.166$, which corresponds to the average adaptive guidance value of the full model. This configuration achieves 11.93~dB SI-SDRi and 0.142 ESTOIi, indicating the contribution of uncertainty-aware training with its use during sampling. To isolate the effect of adaptive guidance, we evaluate the fully trained model at inference with a fixed $\lambda=0.166$, which achieves 12.98~dB SI-SDRi and 0.172 ESTOIi. Lastly, training with the PIT MSE objective on target-active frames without boundary-aware weighting and uncertainty modeling achieves 10.52~dB SI-SDRi and 0.138 ESTOIi.

\begin{table}[!t]
    \centering
    \caption{Ablation study of MOUSSE components.}
    \label{tab:ablation}
    \resizebox{0.98\columnwidth}{!}{%
    \begin{tabular}{l|ccc}
        \toprule[1.3pt]
        Configuration & SDRi & SI-SDRi & ESTOIi \\
        \Xhline{2\arrayrulewidth}
        \textbf{Full model} & 13.20 & 13.18 & 0.175 \\
        \hline
        -- Boundary-aware weighting & 12.11 & 12.10 & 0.164 \\
        -- NLL loss ($\lambda=0.166$) & 11.95 & 11.93 & 0.142 \\
        Fixed guidance ($\lambda=0.166$) & 13.00 & 12.98 & 0.172 \\
        PIT MSE training & 10.53 & 10.52 & 0.138 \\
        \bottomrule[1.3pt]
    \end{tabular}%
    }
\end{table}

\subsection{Uncertainty Analysis}

We further analyze how the predicted scale varies across mixture conditions and whether it reflects model prediction error. Table~\ref{tab:uncertainty} summarizes the mean conditional predicted scale $\sigma_{\mathrm{cond}}$, averaged over sampling steps and active frames, across language compositions and overlap ratios.
As audio overlap increases from 0\% to 100\%, the predicted scale rises from 0.149 to 0.198, from 0.137 to 0.217, and from 0.131 to 0.221 for two- to four-speaker mixtures, respectively. This consistent increase indicates that the estimator assigns larger scales under higher overlap ratios. Across language compositions, English-only mixtures achieve the highest SI-SDRi, outperforming Korean-only mixtures by 3.98 dB. In contrast, multilingual mixtures demonstrate lower Si-SDRi, which suggests that cross-lingual overlap introduces additional challenges for speech separation.

\begin{figure}[t!]
    \centering
    \begin{subfigure}[b]{0.48\linewidth} 
        \centering
        \includegraphics[width=\textwidth]{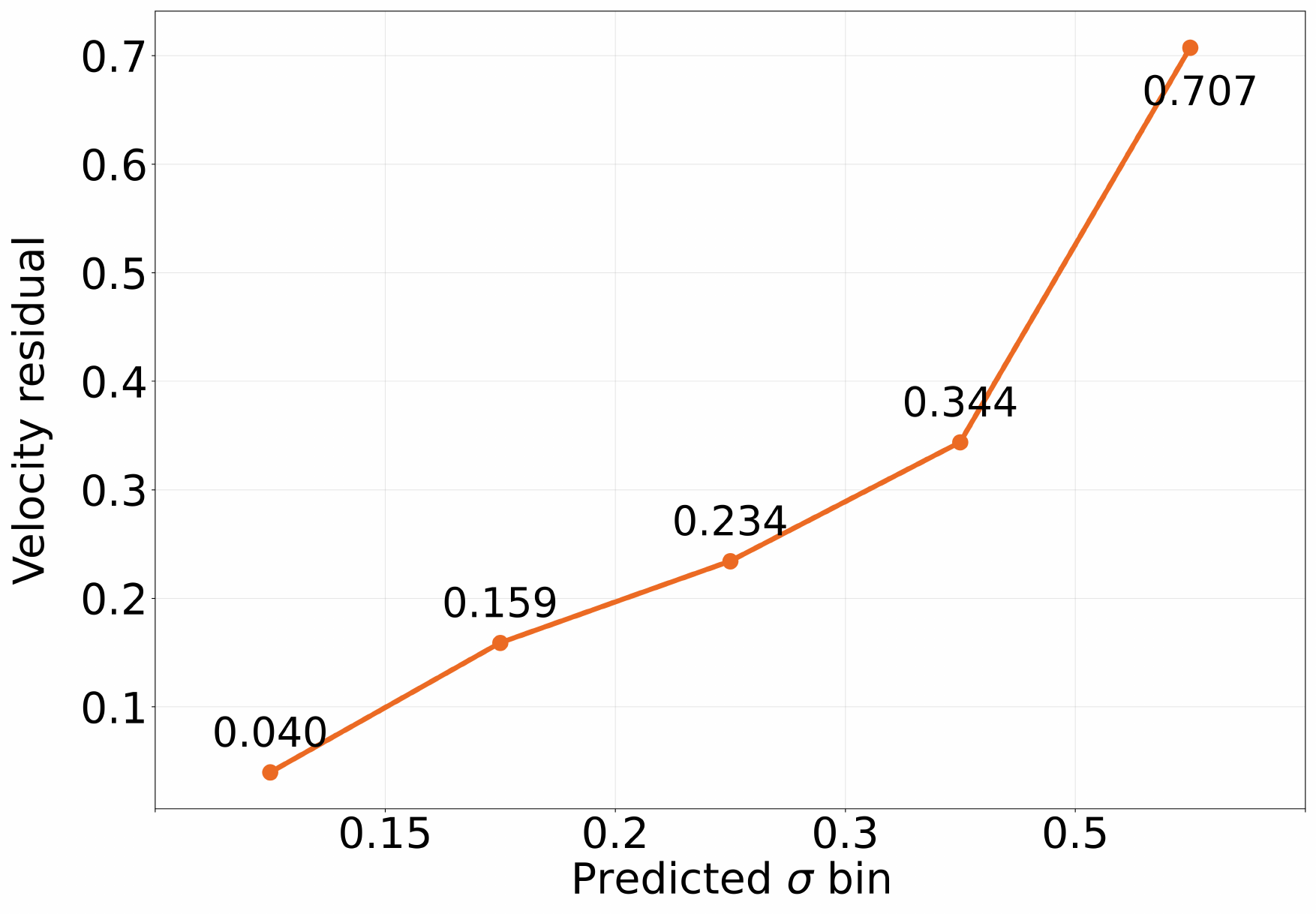}
        \caption{Residual by predicted scale.}
        \label{fig:residual}
    \end{subfigure}
    \begin{subfigure}[b]{0.48\linewidth}
        \centering
        \includegraphics[width=\textwidth]{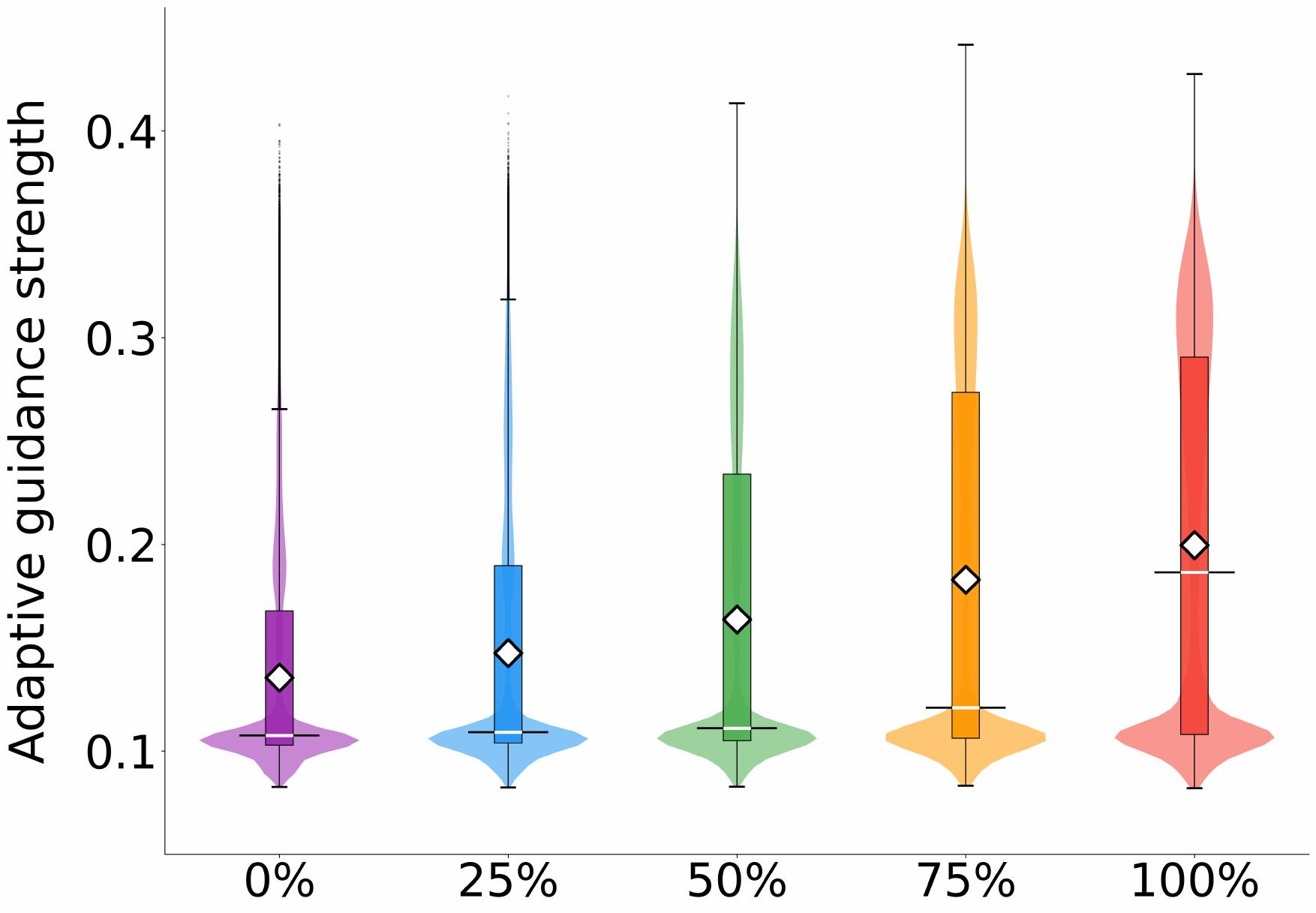}
        \caption{Guidance by overlap ratio.}
        \label{fig:guidance}
    \end{subfigure}
    \caption{Predictive uncertainty and adaptive guidance analysis.}
    \label{fig:uncertainty-analysis}
\end{figure}

To further examine the frame-level relationship between predicted scale and prediction error, Figure~\ref{fig:residual} groups active frames by predicted-scale range and shows that the mean velocity residual increases consistently with predicted scale. A Spearman correlation of $\rho=0.884$, computed across active frames between predicted scale and velocity residual, supports this relationship, indicating that larger predicted scales are associated with larger flow-prediction errors. These results suggest that the predicted scale reflects local variations in flow-prediction error and can therefore serve as a control signal for adaptive guidance.

Moreover, guidance distribution shifts toward larger values as overlap increases, while substantial variation remains within each condition (Fig.~\ref{fig:guidance}). The concentration near low $\lambda^\star$ values reflects the large number of frames with relatively small predicted scales, while higher-overlap conditions contain more frames with larger guidance strengths. This variation thus validates using predicted scale to adjust guidance strength for each frame.

\begin{table}[t!]
    \caption{Mean conditional predicted scale $\sigma_{\mathrm{cond}}$ and SI-SDRi across language compositions and overlap ratios.}
    \label{tab:uncertainty}
    \setlength{\tabcolsep}{6pt}
    \centering\resizebox{0.88\columnwidth}{!}{
    \begin{tabular}{c|l|ccc||c}
        \toprule[1.2pt]
         &  & 2 spk & 3 spk & 4 spk & SI-SDRi \\
        \Xhline{2\arrayrulewidth}
        \multirow{3}{*}{Lang.}
        & Eng.   & 0.166 & 0.168 & 0.172 & 14.79 \\
        & Kor.   & 0.193 & 0.193 & 0.181 & 10.81 \\
        & Mixed  & 0.168 & 0.171 & 0.170 & 13.89 \\
        \hline
        \multirow{5}{*}{\makecell{Overlap\\Ratio}}
        & 0\%    & 0.149 & 0.137 & 0.131 & 13.77 \\
        & 25\%   & 0.166 & 0.152 & 0.142 & 13.95 \\
        & 50\%   & 0.181 & 0.175 & 0.162 & 13.21 \\
        & 75\%   & 0.197 & 0.200 & 0.190 & 11.54 \\
        & 100\%  & 0.198 & 0.217 & 0.221 & 13.76 \\
        \bottomrule[1.2pt]
    \end{tabular}
    }
\end{table}

\section{Conclusion}
\label{sec:conclusion}

In this paper, we have presented MuSSeD, an English--Korean speech separation benchmark that covers 2--4 speakers and multiple overlap conditions, and MOUSSE, a multilingual flow matching framework with predictive uncertainty. MOUSSE incorporates boundary-aware training and frame-wise predictive uncertainty, achieving 13.18~dB SI-SDRi and 0.175 ESTOIi. Ablation results show that boundary-aware weighting and uncertainty-aware training provide larger improvements, while adaptive guidance provides a relatively smaller improvement over fixed guidance. Future work will extend evaluation to additional languages and real conversational settings with background noise.

\section{ACKNOWLEDGMENTS}
\label{sec:ack}

This research was partly supported by Basic Science Research Program through the National Research Foundation of Korea(NRF) funded by the Ministry of Education(RS-2022-NR070870), by Institute of Information \& communications Technology Planning \& Evaluation (IITP) grant funded by the Korea government(MSIT) (No.RS-2019-II191906, Artificial Intelligence Graduate School Program(POSTECH)), and by the Gyeongsangbuk-do RISE(Regional Innovation System \& Education) project(B0080527002599). This work was also supported in part by the Natural Sciences and Engineering Research Council (NSERC) of Canada and in part by the AI2050 program at Schmidt Sciences. Finally, we are grateful for the support from IVADO and the Canada First Research Excellence Fund.

\section{Compliance with Ethical Standards}
This is a modeling study for which no ethical approval was required.

\bibliographystyle{IEEEbib}
\bibliography{strings,refs}

\debug{
\section{Debug Statistics}
\begin{center}
    Unknowns - \arabic{unknowncounter} \\
    Comments - \arabic{commentcounter} \\
    Need charts - \arabic{needchartcounter}
\end{center}
}

\end{document}